\documentclass[aps,prl,twocolumn, showpacs,
  superscriptaddress]{revtex4}

\usepackage{graphicx} \usepackage{color} \usepackage{natbib}
\usepackage{epsfig}

\def\spacelapse{\sigma}
\def\spaceshift{\omega}
\def\xitwo{\xi_2}
\def\xione{\xi_1}

\usepackage{amsmath,amssymb,amsfonts}
\usepackage{url}
\begin{document}

\title{Trumpet Slices in Kerr Spacetimes}

\author{Kenneth A. Dennison}
\affiliation{Department of Physics and Astronomy, Bowdoin College, Brunswick, ME 04011, USA}

\author{Thomas W. Baumgarte}
\affiliation{Department of Physics and Astronomy, Bowdoin College, Brunswick, ME 04011, USA}

\author{Pedro J. Montero}
{\affiliation{Max-Planck-Institut f\"ur
  Astrophysik, Karl-Schwarzschild-Str.~1, 85748, Garching, Germany}

\begin{abstract}
We introduce a new time-independent family of analytical coordinate systems for the Kerr spacetime representing rotating black holes.  We also propose a (2+1)+1 formalism for the characterization of trumpet geometries.  Applying this formalism to our new family of coordinate systems we identify, for the first time, analytical and stationary trumpet slices for general rotating black holes, even for charged black holes in the presence of a cosmological constant.   We present results for metric functions in this slicing and analyze the geometry of the rotating trumpet surface.
\end{abstract}


\maketitle

Many numerical relativity simulations adopt a 3+1 decomposition in which the four-dimensional spacetime is split into a foliation of three-dimensional spatial slices. In the context of such a 3+1 decomposition the coordinate conditions are imposed with the help of a lapse function and a shift vector.  A particular successful choice of coordinates for the evolution of black-hole spacetimes are so-called moving-puncture coordinates (see, e.g., \cite{CamLMZ06,BakCCKM06a} as well as numerous later simulations; see also \cite{BauS10} for a pedagogical introduction).   When evolved with moving-puncture coordinates, black-hole spacetimes settle down to a foliation in which the spatial slices take on a {\em trumpet} geometry.   Trumpet slices end on a two-dimensional {\em trumpet surface} that is embedded in the spatial slices and encloses the spacetime singularity.  The slices therefore avoid spacetime singularities, and allow numerical simulations of black-hole spacetimes without special treatment of the black holes.

The geometric properties of static trumpet slices of (nonrotating) Schwarzschild black holes \cite{Sch16} are well understood (see, e.g., \cite{HanHPBO06,BauN07,HanHOBO08,Bro08,Bru09}).  On the trumpet surface the lapse vanishes (marking the boundary of the spatial slice), the surface has a finite and non-zero proper area (ensuring that the surface is removed from the spacetime singularity) and it is an infinite proper distance away from all points outside the trumpet surface itself (so that the rest of spacetime is not affected by the presence of the coordinate singularity).   An embedding diagram, which resembles a trumpet and gives these slices their name, is shown, for example, in Fig.~2 of \cite{HanHOBO08}.  Understanding these properties has been very helpful in both interpreting and guiding numerical simulations.  While the gauge conditions used in many numerical relativity simulations result in trumpet slices that cannot be given in completely analytical form, we have recently presented a different but completely analytical family of trumpet slices of the Schwarzschild spacetime in \cite{DenB14}.

Generic black holes, however, rotate, and generic numerical relativity simulations result in Kerr black holes \cite{Ker63}.  Evidently it would therefore be desirable to gain a better understanding of the geometric properties of trumpet slices of the Kerr spacetime.  While this has been recognized as an interesting and important problem, it appears difficult to generalize analytical results for those trumpet slices realized for the gauge conditions used in many numerical simulations (see, e.g., \cite{DieB14} for a numerical study; see also 
\cite{DaiC09,ImmB09,HanHO09,Cle10,Bau12,ChrM12} for approaches to constructing trumpet initial data for rotating black holes).  In this paper, we instead adopt the procedure of \cite{LinS13} to generalize the above-mentioned family of analytical trumpet slices \cite{DenB14} to rotating black holes.  We thereby introduce a new time-independent analytical coordinate system for the Kerr spacetime.  

It is quite easy to verify that spherically symmetric slices of the Schwarzschild spacetime can simultaneously have all three properties of a trumpet surface proposed above (vanishing lapse, finite proper area, and infinite proper distance from any point off the surface).   In the absence of spherical symmetry it is not only more complicated to evaluate these properties; a priori it is not even clear whether all three conditions can be met simultaneously.  Below we propose a (2+1)+1 formalism for the characterization of trumpet slices in axisymmetric spacetimes, and we demonstrate that slices of constant coordinate time in our new coordinate system for Kerr spacetimes do indeed meet these criteria.  With the exception of extreme Kerr black holes, for which surfaces of constant Boyer-Lindquist time form trumpet slices (see, e.g., \cite{DaiC11}), our solutions represent, to the best of our knowledge, the first analytical examples of stationary trumpet slices in general rotating black holes. 

We start with a 3+1 decomposition of a stationary, axisymmetric spacetime ${\mathcal M}$.   We will assume below that the spacetime metric $g_{ab}$ is given in terms of spherical polar coordinates $t$, $R$, $\theta$, and $\phi$, but independent of $t$ and $\phi$.  We then introduce a foliation $\Sigma$ of ${\mathcal M}$ that is formed by level-surfaces of the coordinate time $t$;  the spacetime metric $g_{ab}$ can then be written in the form
\begin{equation} \label{threeplusone}
g_{ab} = \left( \begin{array}{cc}
- \alpha^2 + \beta_i \beta^i  & \beta_i \\
\beta_j	& \gamma_{ij} 
\end{array}
\right)
\end{equation}
where $\alpha$ is the lapse function, $\beta^i$ the shift vector, and $\gamma_{ab} \equiv g_{ab} + n_a n_b$ the spatial metric induced by $g_{ab}$ on the spatial slice.  Indices $a, b, \ldots$ run over spacetime indices, while indices $i, j, \ldots$ run over spatial indices only, and
\begin{equation}
n_a = (- \alpha,0,0,0)
\end{equation}
is the future-pointing normal on the slices $\Sigma$.  The proper time $\tau$ as measured by normal observers advances according to $d\tau = \alpha dt$.  We also note that the determinant $g$ of the spacetime metric is given by
\begin{equation} \label{spacetimedet}
- g = \alpha^2 \gamma,
\end{equation}
where $\gamma \equiv \det(\gamma_{ij})$.

We now perform an analogous 2+1 decomposition of the spatial slices.  We consider axisymmetric, closed hypersurfaces $S$ of the spatial slices $\Sigma$, centered on the origin, that can be represented as level surfaces of a (potentially) new radial coordinate $\bar R = \bar R(R,\theta)$.  In complete analogy to the above, we can then write the spatial metric $\gamma_{ij}$, in the new barred coordinates, in the form
\begin{equation} \label{twoplusone}
\gamma_{\bar \imath \bar \jmath} = \left( \begin{array}{cc}
 \spacelapse^2 + \spaceshift_A \spaceshift^A  & \spaceshift_A \\
\spaceshift_B	& h_{AB} 
\end{array}
\right),
\end{equation}
where $\spacelapse$ and $\spaceshift^A$ play the same roles as $\alpha$ and $\beta^i$ above, and where $h_{\bar \imath \bar \jmath} \equiv \gamma_{\bar \imath \bar \jmath} - s_{\bar \imath} s_{\bar \jmath}$ is the surface metric induced by $\gamma_{\bar \imath \bar \jmath}$ on $S$.  Indices $A$, $B$ \ldots run over angular indices only,  and
\begin{equation}
s_{\bar \imath} = (\spacelapse,0,0)
\end{equation}
is the outward-pointing normal on the surfaces $S$.  The proper distance between two surfaces, measured along the normal, advances according to 
\begin{equation} \label{prop_dist}
dl = \spacelapse d\bar R.
\end{equation}
In analogy to (\ref{spacetimedet}), the determinant $\gamma$ may be expressed as
\begin{equation} \label{spacedet}
\gamma =  J^2 \bar \gamma = J^2 \spacelapse^2 h,
\end{equation}
where $\bar \gamma \equiv \det(\gamma_{\bar \imath \bar \jmath})$, $h \equiv \det(h_{AB})$ and where we assume the Jacobian of the transformation from the unbarred to the barred spatial coordinates $J \equiv \det(\partial x^{\bar \imath}/\partial x^j)$ to be finite and non-zero.  Combining (\ref{spacetimedet}) with (\ref{spacedet}) we also have
\begin{equation} \label{det}
- \hat g = J^2 \alpha^2 \spacelapse^2 \hat h
\end{equation}
where $\hat g \equiv g/\sin^2 \theta$ and $\hat h \equiv h/\sin^2\theta$.

We can now characterize a trumpet surface at, say, $\bar R = \bar R_0$ as follows.  We require that this surface surround all spacetime singularities and hence have finite (and non-zero) proper area; we will therefore assume that $\hat h$ be finite (and non-zero) at $\bar R = \bar R_0$.  We next require that the surface have an infinite proper distance from any point $\bar R > \bar R_0$; according to (\ref{prop_dist}) this means that $\spacelapse^{-1}$ must have (at least) a single root at $\bar R_0$,
\begin{equation}
\spacelapse \propto (\bar R - \bar R_0)^{-n}
\end{equation}
with $n \geq 1$.  As long as $\hat g$ remains finite at $\bar R_0$, relation (\ref{det}) then shows that the lapse automatically has at least a single root, marking the boundary of the spatial slice.  In fact, these arguments show that, as long as $\hat g$ remains finite and non-zero at $\bar R_0$, a trumpet surface can be identified as a closed surface with finite $\hat h$ on which the lapse $\alpha$ takes at least a single root.

In \cite{DenB14} we presented an analytical family of trumpet slices for Schwarzschild black holes, parameterized by the areal radius of the trumpet surface $0 \leq R_0 \leq M$.  The family contains, as a special member, Painlev\'e-Gullstrand coordinates \cite{Pai21,Gul22} for $R_0 = 0$ (for which the trumpet disappears).  Several authors (including \cite{Dor00,ZhaZ05,Nat09,LinS13}) have suggested procedures that generalize Painlev\'e-Gullstrand coordinates for Kerr black holes.  We now adopt the procedure of \cite{LinS13} to generalize the entire family of trumpet slices for rotating black holes.  As discussed in \cite{LinS13}, we can transform from Boyer-Lindquist coordinates \cite{BoyL67} ($t_{BL}$, $R_{BL}$, $\theta_{BL}$, $\phi_{BL}$) to generalized Painlev\'e-Gullstrand coordinates ($t$, $R$, $\theta$, $\phi$) by defining 
\begin{equation}
dt_{BL} = dt-\frac{(R^2+a^2)\sqrt{f^{2}-(R^2-2MR+a^2)}}{(R^2-2MR+a^2)f}dR,
\end{equation}
and
\begin{equation}
d\phi_{BL} = d\phi - a\frac{\sqrt{f^{2}-(R^2-2MR+a^2)}}{(R^2-2MR+a^2)f}dR,
\end{equation}
as well as $dR_{BL} = dR$ and $d\theta_{BL} = d\theta$, where $f\equiv f(R)$ is an arbitrary function.  Choosing $f(R) = R - R_0$ we arrive at the line element
\begin{eqnarray}
\label{Kerr_metric}
ds^2 &=& -\frac{\rho^2-2 M R}{\rho^{2}}dt^{2}\nonumber\\&&+2\frac{\sqrt{R_{0}^{2}+2R(M-R_{0})-a^{2}}}{R-R_{0}}dtdR\nonumber\\&&-\frac{4 a M R\sin^{2}\theta}{\rho^{2}}dtd\phi+\frac{\rho^{2}}{(R-R_{0})^{2}}dR^2\nonumber\\&&-2a\frac{\sqrt{R_{0}^{2}+2R(M-R_{0})-a^{2}}}{R-R_{0}}\sin^{2}\theta dRd\phi\nonumber\\&&+\rho^{2}d\theta^2+
\frac{\sin^{2}\theta}{\rho^{2}} \, \xione d\phi^{2}.
\end{eqnarray}
Here $M$ is the black hole's mass, $aM$ its angular momentum, $R_0$ is a -- so far -- arbitrary constant, and we have defined
\begin{equation}
\rho \equiv \sqrt{R^{2}+a^{2}\cos^{2}\theta}
\end{equation}
as well as
\begin{equation}
\xione \equiv  \rho^{2} ( R^2 + a^2)  +2 a^{2}M R\sin^{2}\theta.
\end{equation}
We have verified that this solution satisfies Einstein's equations.  In the limit of zero rotation, $a=0$,  we recover the expressions of \cite{DenB14} for the Schwarzschild spacetime; for extreme Kerr, $a = M$, we recover the metric in Boyer-Lindquist coordinates, provided we choose $R_0 = M$.

It is now straightforward to verify that slices of constant coordinate time are trumpet slices.  We first compute
\begin{equation} \label{g_det}
-\hat g = \rho^{4}
\end{equation}
which is non-zero and finite as long as $\rho$ is, so that the arguments following eq.~(\ref{det}) apply.  We then perform the 3+1 decomposition (\ref{threeplusone}) and identify the lapse
\begin{equation} \label{lapse}
\alpha = \frac{\rho(R-R_{0})}{\xitwo^{1/2}},
\end{equation}
where we have abbreviated
\begin{equation}
\xitwo = (R^{2}+a^{2})^{2}-a^{2}(R-R_{0})^{2}\sin^{2}\theta,
\end{equation}
as well as the spatial metric $\gamma_{ij}=g_{ij}$ \footnote{We note that the ADM mass can be defined only for $R_0 = M$; for all other choices the components of the spacetime metric do not fall off sufficiently fast asymptotically.  Since our data are stationary, however, the Komar mass can be evaluated for all values of $R_0$ (see also the discussion in [3]).}.   For completeness we also list the non-zero components of the shift
\begin{subequations} \label{shift}
\begin{equation}
\label{shiftR}
\beta^{R} = \frac{R^{2}+a^{2}}{\rho^{2}} \, \alpha^{2}g_{tR}
\end{equation}
and
\begin{equation}
\label{shiftphi}
\beta^{\phi} = -a \frac{ R^{2}+a^{2} -(R-R_{0})^{2}}{\rho^{2}(R-R_{0})^{2}}\,\alpha^{2}.
\end{equation}
\end{subequations}
Evidently, the lapse (\ref{lapse}) has a single root in $R$ at $R = R_{0}$, making this coordinate-sphere a candidate for a trumpet surface.  We therefore do not need to transform to a new radial coordinate $\bar R$, and instead may apply the 2+1 decomposition (\ref{twoplusone}) directly to surfaces of constant $R$.  Dropping the bars in the above expressions we identify
\begin{equation}
h_{AB} = \left( \begin{array}{cc}
\gamma_{\theta\theta}  &  0\\
0 &  \gamma_{\phi\phi}
\end{array}
\right).
\end{equation}
The rescaled determinant of this metric
\begin{equation} \label{h_det}
\hat h = \gamma_{\theta\theta}\gamma_{\phi\phi}/\sin^{2}\theta = \xione
\end{equation}
is finite and non-zero at $R = R_0$, as we required above for a trumpet surface.   Eq.~(\ref{det}) now implies automatically that this surface is an infinite proper distance away from all points with radii $R > R_0$.  To verify this, we identify
\begin{equation} \label{sigma}
\sigma^{2}=\gamma_{RR}-\frac{\gamma_{R\phi}^{2}}{\gamma_{\phi\phi}}=\frac{\rho^{2}}{(R-R_{0})^{2}}\frac{\xitwo}{\xione}
\end{equation}
from (\ref{twoplusone}), so that the integral (\ref{prop_dist}) indeed diverges at $R = R_0$.   We can also insert eqs.~(\ref{g_det}), (\ref{lapse}), (\ref{h_det}) and (\ref{sigma}) into (\ref{det}) to verify that this equation is satisfied with $J = 1$.   This completes the identification of $R = R_0$ surfaces as trumpet surfaces in the Kerr spacetime.  


\begin{figure}
\centering
\includegraphics[width=0.4\textwidth]{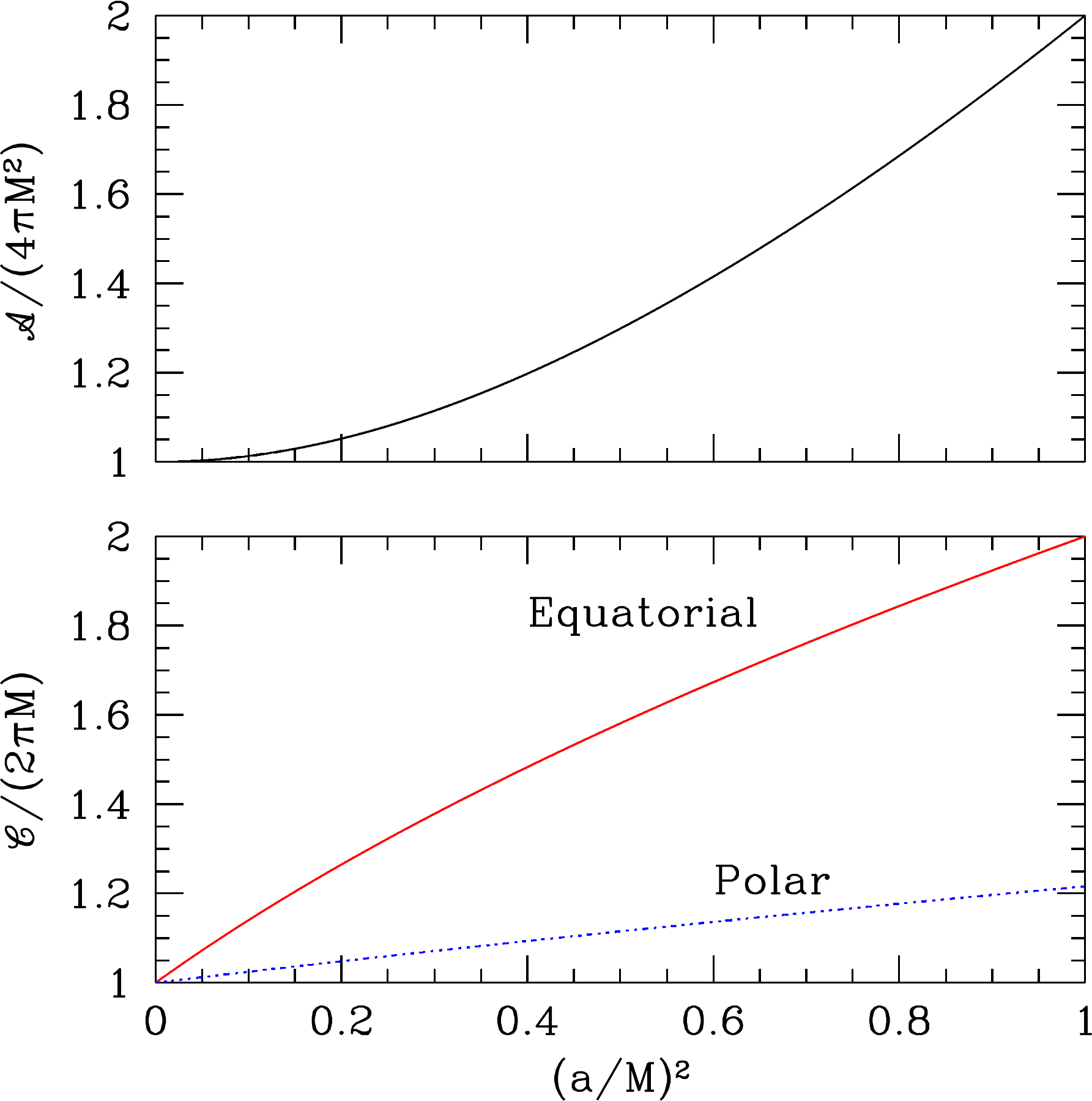}
\caption{The proper area ${\mathcal A}$ of the trumpet surface (top panel) as well as the proper circumferences ${\mathcal C}$ (bottom panel) of the trumpet surface as a function of the squared spin parameter $(a/M)^2$, for $R_0 = M$.  The equatorial circumference is measured along the equator at $\theta = \pi/2$, while the polar circumference is measured at constant $\phi$.}
\label{fig1}
\end{figure}


\begin{figure}
\centering
\includegraphics[width=0.4\textwidth]{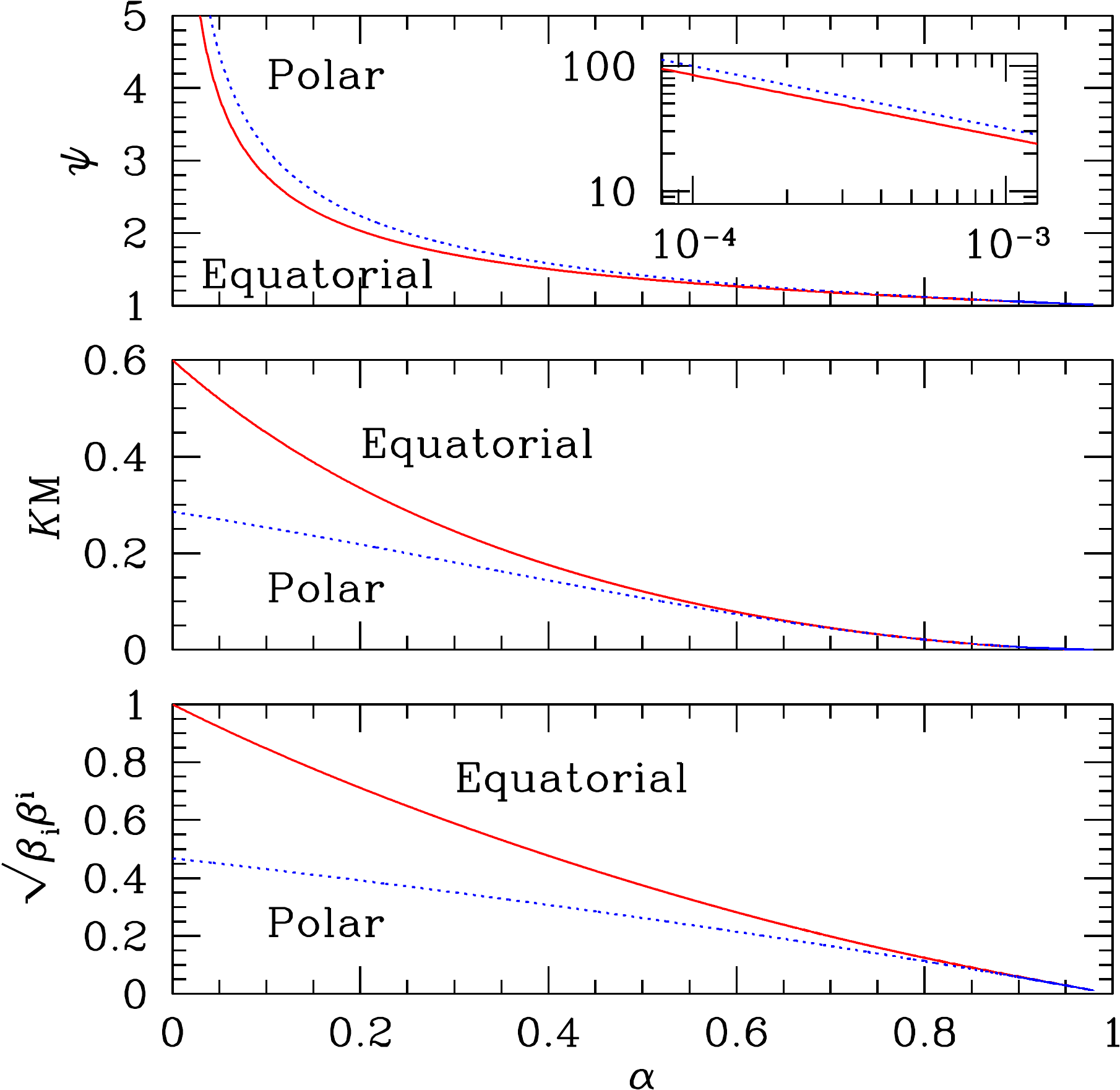}
\caption{The conformal factor $\psi$ (\ref{conf_factor}), the trace $K$ of the extrinsic curvature (\ref{traceKij}) and the magnitude of the shift $\sqrt{\beta_{i}\beta^{i}}$ (\ref{shift}) as a function of the lapse $\alpha$ (\ref{lapse}) (which, unlike the radius, is invariant under spatial coordinate transformations).   Vanishing lapse $\alpha = 0$ corresponds to the trumpet surface, while $\alpha = 1$ is spatial infinity.  All graphs are shown for $a = 0.8 M$ and $R_0 = M$.  In each case the solid (red online) curve shows the relationship in the equatorial plane, while the dotted (blue online) curve shows the relationship in the polar direction.  The inset in the top panel shows the conformal factor near the trumpet surface on a log-log scale.  On this graph the slope of both lines is indistinguishable from the slope of
$\alpha^{-1/2} \propto (R - M)^{-1/2} = r^{-1/2}$, where the proportionality follows from eqs.~(\ref{lapse}) and (\ref{conf_factor}), and where 
$r$ is a quasi-isotropic radius (see below).}
\label{fig2}
\end{figure}

For $g_{tR}$ to be real, the free parameter $R_0$ should be chosen within the limits $M-\sqrt{M^{2}-a^{2}}\leq R_{0}\leq M$, meaning that the trumpet surface is always between the inner and outer horizon of the Kerr black hole.  The only choice of $R_0$ that can be used for all values of $0 \leq a^2 \leq M^2$ is $R_0 = M$, which further simplifies some of the above expressions (see also \cite{DenB14}).  In the Figures we show some results for this choice.  In particular, we show the proper area (top panel) as well as both the equatorial and polar circumferences (bottom panel) as functions of $(a/M)^2$ in Fig.~\ref{fig1}.   In Fig.~\ref{fig2} we show the magnitude of the shift (\ref{shift}), the trace $K$ of the extrinsic curvature $K_{ij}$, 
\begin{equation} \label{traceKij}
K = \gamma^{ij}K_{ij} = \frac{\sqrt{M^2-a^2}}{\rho^2}\partial_{R}\left(\frac{(R^2 + a^2)\alpha}{R-M}\right),
\end{equation}   
and a conformal factor $\psi$ which, for our purposes here, we define as
\begin{equation} \label{conf_factor}
\psi \equiv \left(\frac{\gamma}{(R-M)^4 \sin^{2}\theta}\right)^{1/12} = \frac{\rho^{1/6}\xi_{2}^{1/12}}{\sqrt{R-M}},
\end{equation} 
as a function of the lapse (\ref{lapse}) for $a=0.8M$.  

The above results can be extended to Kerr-Newman-de Sitter black holes, i.e.~rotating charged black holes \cite{NewCCEPT65} in the presence of a cosmological constant $\Lambda$ \cite{Car68} with $\Lambda>-3/a^{2}$ for nonzero $a$.
Defining
\begin{equation}
\Delta \equiv R^2 - 2 M R + a^2 -\frac{\Lambda R^2 (R^2+a^2)}{3}+Q^2,
\end{equation}
as well as
\begin{equation}
\Xi \equiv 1+\frac{\Lambda a^{2}}{3}
\mbox{~~~~and~~~~}
\Xi_{\theta} \equiv 1+\frac{\Lambda a^{2}}{3}\cos^{2}\theta,
\end{equation}
we find that the line element is
\begin{widetext}
\begin{eqnarray}
\label{Kerr_Newman_deSitter_metric}
ds^2 &=& -\frac{\Delta-a^{2}\Xi_{\theta}\sin^{2}\theta}{\Xi^{2}\rho^{2}}dt^{2}+2\frac{\sqrt{(R-R_{0})^2-\Delta}}{\Xi(R-R_{0})}dtdR-2\frac{a \left(\Lambda\rho^{2}(R^{2}+a^{2})/3 + 2 M R-Q^2\right)\sin^{2}\theta}{\Xi^{2}\rho^{2}}dtd\phi\nonumber\\&&+\frac{\rho^{2}}{(R-R_{0})^{2}}dR^2-2a \frac{\sqrt{(R-R_{0})^2-\Delta}}{\Xi(R-R_{0})}\sin^{2}\theta dRd\phi+\frac{\rho^{2}}{\Xi_{\theta}}d\theta^2\nonumber\\&&+\frac{\sin^{2}\theta}{\Xi^2\rho^{2}}\left(\left(R^2 + a^2\right)\left(R^2 + a^2 \left(1+\frac{\Lambda\rho^2}{3}\right)\right)-a^2 \left(R^2-2 M R +a^2 + Q^2\right)\sin^{2}\theta\right)d\phi^2.  
\end{eqnarray}
\end{widetext}  
For $\Lambda = 0$ and $Q=0$, the Kerr-Newman-de Sitter metric (\ref{Kerr_Newman_deSitter_metric}) reduces to the Kerr metric (\ref{Kerr_metric}), while for $a=0$ and $Q=0$ it reduces to an extension of the family of \cite{DenB14} to Schwarzschild-de Sitter spacetimes \cite{Kot18,Wey19,Tre22,Bon14}.  For $a=0$ and $\Lambda = 0$, it reduces to a family of trumpet slicings of the Reissner-Nordstr{\o}m spacetime \cite{Rei16,Nor18}.  As before, slices of constant coordinate time $t$ are trumpet slices, with the trumpet surface at $R = R_0$.

Most numerical simulations adopt quasi-isotropic spatial coordinates, for which the coordinate radius $r$ of the trumpet surface vanishes.   The above solution can be transformed to such a coordinate system very easily with the transformation $r = R - R_0$ (for which the spatial metric becomes isotropic in the limit $a=0$.)  We note, however, that $K$ is not single-valued on the trumpet surface (see also Fig.~\ref{fig2}).  This is one indication that our new coordinate system for Kerr is not well-suited for numerical simulations (see also the discussion in \cite{DenB14}).  It is also not clear, a priori, whether the criteria for trumpet surfaces (vanishing lapse, finite proper area, and infinite proper distance from any point off the surface) are generally compatible with the gauge conditions typically used in numerical relativity simulations of black holes.  The point of this paper, however, is to introduce a (2+1)+1 formalism for the characterization of trumpet surfaces, and to demonstrate analytically that such surfaces do indeed exist in the spacetimes of rotating black holes.  We introduce a surprisingly simple new coordinate system for the Kerr spacetime, and present  the first analytical examples of stationary trumpet slices for general rotating black holes.

\acknowledgments

We would like to thank Beatrice Bonga for drawing our attention to de Sitter spacetimes and for providing her notes on the Schwarzschild-de Sitter spacetime \cite{Bon14}.  Some calculations were assisted by Mathematica \cite{Wol12} and the RGTC package \cite{Rgt13} as well as the Sage \cite{Sag12} package SageManifolds \cite{Sag14}.  This work was supported in part by NSF grants PHY-1063240 and PHY-1402780 to Bowdoin College, and the Deutsche Forschungsgemeinschaft (DFG) through its Transregional Center SFB/TR7 ``Gravitational Wave Astronomy''.


\end{document}